# Quantum Communication – Nicolas Gisin, University of Geneva, Switzerland

**Status** – Quantum Communication (QC) is the art of transferring an unknown quantum state from one location, Alice, to a distant one, Bob [1]. This is a non-trivial task because of the quantum no-cloning theorem which prevents one from merely using only classical means. On the one hand, QC is fascinating because it allows one to distribute entanglement over large distances and to thus establish so-called non-local correlations as witnessed by violations of Bell inequalities, i.e. correlations that can't be explained with only local variables that propagate contiguously through space. On the other side, QC has enormous potential applications. The best known one is Quantum Key Distribution (QKD), the use of entanglement to guarantee the secrecy of keys ready to use in cryptographic applications. Other examples of possible applications are processes whose randomness is guaranteed by the violation of some Bell inequality. Indeed, it is impossible to prove that a given sequence of bits is random, but it is possible to prove that some processes are random and it is known that random processes produce random sequences of bits. Hence, randomness moves from mathematics to physics.

In QC one needs to master the photon sources, the encoding of some quantum information (quantum bit, or in short qubit), the propagation of the photons while preserving the quantum information encoded in the photons (e.g. polarization or time-bin qubits), possibly teleport that information and/or store it in a quantum memory, and eventually detect the photon after an analyzer that allows one to recover the quantum information. Additionally, QC evolves from the mere point-to-point configuration to complex networks. The latter require, on the fundamental side, better understanding of multi-partite entanglement and non-locality and, on the experimental side, a global system approach with good synchronization, network control, and recovery processes.

**Current and Future Challenges** – QC faces many challenges that range from industrializing quantum technologies to conceptual questions in theoretical physics, spanning various fields of physics, from optics to material science.

Sources: Besides the relatively simple case of point-to-point QKD all QC tasks require sources of entangled photons. Today such sources are based on spontaneous parametric down-conversion, hence are probabilistic. Consequently, in order to avoid multiple pairs of photons, the probability of a single pair is maintained quite low, typically 1% to 0.1%. Moreover, coupling losses between the nonlinear crystals in which the photons are created and the optical fiber are critical. A

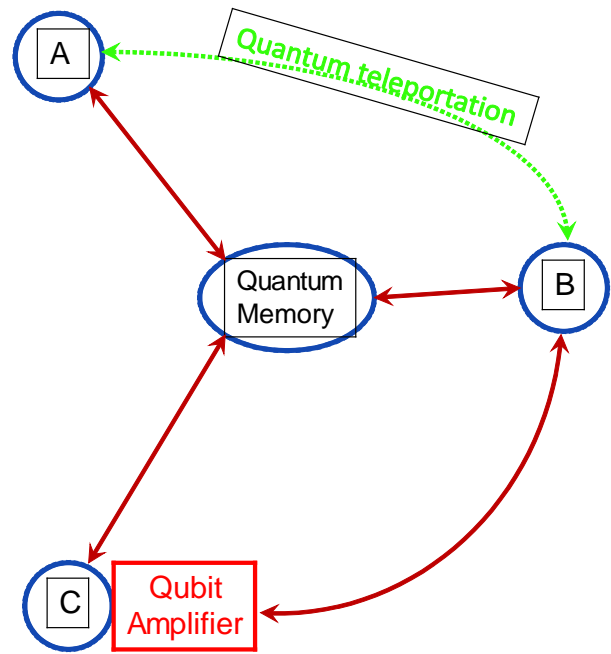

grand challenge is to develop handy and deterministic sources of entangled photons.

The quantum channels could be either free space or an optical fiber. The former is mostly convenient for earth-to-satellite and satellite-to-satellite QC; there, the main challenges are the size and weight of the telescopes [2]. For the fibers, the main challenge is losses. Today's best fibers have losses as low as 0.16 dB/km, i.e. after 20 km almost half the photons are still present [3].

Quantum teleportation: This fascinating process allows one to transfer a quantum states (i.e. quantum information) using pre-established entanglement as a channel. The quantum state doesn't follow any trajectory in space, but is "teleported" from here to there. In addition to the distribution of entanglement, teleportation requires joint measurements, another quantum feature non-existing in classical physics. Joint measurements allow one to measure relative properties between two systems, like "are your polarization states anti-parallel?", without gaining any information about the individual properties [4]. Hence, the individual properties don't get disturbed, but the correlations between the properties of the two systems get "quantum correlated", i.e. entangled. In practice, two photons get mixed on a beam-splitter. This works, but requires high stability as the two photons must be indistinguishable in all parameters. In particular timing issues are serious, especially when the photons travel long distances before meeting on the beam-splitter.

Moreover, the process is probabilistic and works at best half the time (but one knows when it worked) [5]. A grand challenge is to greatly improve joint measurements. This requires, probably, to transfer of the photonic quantum states in solids and that the joint measurement is carried out on the degrees of freedom that code the quantum state in the solid, e.g. between two spins. Also, from a purely theoretical standpoint, better understanding of joint measurements, like the abstract formulation of non-local correlations, is much in demand.

Heralded probabilistic qubit amplifier should also be mentioned. This process, inspired by teleportation, allows one to increase the probability of the presence of a photon without perturbing the qubit it encodes [6]. The challenge is to demonstrate such an amplifier over a long distance (>10 km).

Detectors: QC is mostly done using individual photons, hence the requirement of excellent single-photon detectors. Though it should be said that one can also use so-called continuous variables, e.g. squeezed light pulses and homodyne detection systems [7]. Recently, single-photon detectors have made huge progress, with efficiencies increasing in a couple of years from 10-20% to 80-90%, thanks to superconducting detection systems [8]. At the same time, the jitter reduced to below 100ps. Still, there are several grand challenges:

- simpler/cheaper (in particular higher temperatures, at least 3K, possibly high-Tc superconductors).

- photon-number resolving detectors, up to a few tens of photons. This fascinating perspective makes sense, however, only if the detection efficiency is at least 95%.

- and 99-100% efficiency. This seems feasible.

Quantum Memories allow one to convert reversibly and on demand photonic quantum states into and out of some atomic system with long coherence times [9]. They are needed to synchronize complex quantum networks. *They could also turn probabilistic sources into a quasi-deterministic one, provided their efficiency is high enough.* Today, quantum memories are still in the labs. Although all parameters like storage time, efficiency, fidelity, bandwidth and muti-mode capacity have been satisfactorily demonstrated, each demonstration used a different system. Hence, the grand challenge is to develop one system capable of storing 100 photonic qubits (e.g. in 100 time modes) for one second with an efficiency of 90% and fidelity around 95%. This is an enormous challenge. It is unclear whether the best solution will turn out to use single natural or artificial atoms in a cavity or use ensembles of atoms (gas or doped optical crystals). It requires a mix of material science and chemistry (for the crystals), cryogenics (low temperatures seem necessary for long storage times), spectroscopy and radiowave spin echo technics, optics and electronics.

Multi-partite entanglement and non-locality: future complex quantum networks will routinely produce multy-partite entangled states whose full power remains to be discovered. In the case of 2-parties, non-local correlations turned out to be the resource for several remarkable processes, called Device Independent Quantum Information Processes (DIQIP) [10]. Likewise, one can expect quantum correlations in complex networks to open new possibilities that remain to be explored.

QKD: Quantum Key Distribution is the most advanced application of QC. The remaining challenges are mostly industrial and commercial. On the industry side, QKD should become cheaper and much faster (one Gb/s is not unthinkable). On the commercial side, the classical security and crypto communities should be educated, they should understand the potential of quantum physics and that it doesn't render their know-how obsolete, but complements it: quantum doesn't solve all problems, but offers guaranteed randomness and secrecy. Hence, it will always have to be combined with classical security and cryptographic systems.

**Advances in Science and Technology to Meet Challenges** – The most significant challenges mentioned above require a combination of photonics and solid-state devices, be it for the deterministic sources, the detectors and the quantum memories, and probably also improved joint measurements. The basic grand challenge lays thus in integrated hybrid systems. Ideally, all these solid-state devices will be pigtailed to standard telecom optical fibers, hence easy to combine in large networks.

**Concluding Remarks** – Future quantum networks will look similar to today's internet. One will merely buy components and plug them together via optical fibers, all driven and synchronized by a higher order control software. Randomness and secrecy will come for free. Entanglement, non-locality and teleportation will be common and children will get used to them. Applications unthinkable today will proliferate. Such a dream is possible thanks to Quantum Communication.